\documentclass[a4paper,11pt,twoside]{article}
\input{stefmac.sty}
\usepackage{graphics,graphicx,color,latexsym,amssymb,amsmath,amsfonts,amsthm,fancyhdr,verbatim}

\begin{document}

\title{\bf New parallel programming language design: a bridge between brain models and
 multi-core/many-core computers?}
\author{\bf Gheorghe Stefanescu$^1$ and Camelia Chira$^2$\vspace{2mm}\\
\small $^1$~Department of Computer Science, University of Illinois at Urbana-Champaign\snvsp\\
\small 201 N. Goodwin, Urbana, IL 61801, USA\snvsp\\
\small Email: {\tt stefanes@cs.uiuc.edu}\vsp\\
\small $^2$~Department of Computer Science, Babes-Bolyai University\snvsp\\
\small Kogalniceanu 1, Cluj-Napoca, Romania 400084\snvsp\\
\small Email: {\tt cchira@cs.ubbcluj.ro}}
\date{}
\maketitle

${}$\vspace*{-9.5cm}\\\parbox{\textwidth}{\small {\tt To appear in:} ``From Natural Language to Soft
  Computing: New Paradigms in Artificial Intelligence,'', L.A.~Zadeh et.al (Eds.), Editing House of Romanian
  Academy, 2008.}\vspace{7.7cm}

\begin{abstract}

The recurrent theme of this paper is that sequences of long temporal patterns as opposed to sequences of
simple statements are to be fed into computation devices, being them (new proposed) models for brain activity
or multi-core/many-core computers. In such models, parts of these long temporal patterns are already committed
while other are predicted. This combination of matching patterns and making predictions appears as a key
element in producing intelligent processing in brain models and getting efficient speculative execution on
multi-core/many-core computers.  A bridge between these far-apart models of computation could be provided by
appropriate design of massively parallel, interactive programming languages. Agapia is a recently proposed
language of this kind, where user controlled long high-level temporal structures occur at the interaction
interfaces of processes. In this paper Agapia is used to link HTMs brain models with TRIPS
multi-core/many-core architectures.

\end{abstract}

\section{Introduction}

We live in a paradox. On the one hand, recent technological advances suggest the possible transition to
powerful multi-core/many-core computers in the near future. However, in order to be economically viable, such
a major shift must be accompanied with a similar shift in software, where parallel programming should enter
the mainstream of programming practice becoming the rule rather than the exception. Briefly, programs eager
for more computing power are badly needed. On the other hand, there is a critical view that the promises of AI
(Artificial Intelligence) are still to be fulfilled. No matter how much computer power we would have, the
critics say, the advances in key AI areas as image recognition or understanding spoken languages will be
slow. This means that the current AI approach is, according to critics, faulty.

AI already had a major restructuring in the nineties, by adopting the agent-oriented paradigm as a major
research topic.\foo{See \cite{chi07} for a recent presentation, centered on the use of agents for cooperative
  design in a distributed environment.} Jeff Hawkins \cite{on-int} proposes another restructuring of AI by
taking a closer look to the human brain. According to Hawkins, the efficient modelling of the human brain is
of crucial importance if we really want to understand why human beings are so powerful on recognizing images
or performing other similar tasks for which computers are still notoriously weak. Following suggestions
provided by the anatomical structure of the brain, he proposes to use HTMs (Hierarchical Temporal Memories),
hierarchical networks of nodes which work together processing continuous flows of data. While most of the
pieces have been already used by other approaches (neural networks, associative memories, learning machines,
interactive computing models, etc.), Hawkins stresses out the importance of having a unitary, coherent
approach. In his view, the interplay between a systematic study of the brain and a creative design approach
for developing intelligent machines is the solution to the current AI crisis.

Our paper briefly presents HTMs and other key elements of Hawkins model of the brain
\cite{on-int}. Furthermore, it describes the specific features of a particular architecture called TRIPS
(Tera-op, Reliable, Intelligently adaptive Processing System) for multi-core/many-core computers
\cite{computer}. The main contribution of the paper might be the suggestion that Agapia, a recently proposed
language for massively parallel, interactive programs \cite{dr-st07b,pss07}, or similar languages, can
potentially be a bridge between brain models, such as those using HTMs, and multi-core/many-core computers,
particularly those using the TRIPS architectures. To strengthen the suggestion, the paper shows how Agapia
programs for modeling HTMs can be developed and sketches an approach for compiling and running Agapia programs
on TRIPS computers.

\section{HTMs - models for brain activity}

Understanding brain activity was and still is so difficult that even speculative theories on ``how the brain
could work'' are rare. In a recent book ``On Intelligence'' \cite{on-int}, Hawkins proposes a computation
model that may explain the striking differences between the current computers and the brain capabilities,
especially in such areas as visual pattern recognition, understanding spoken language, recognizing and
manipulating objects by touch, etc. Hawkins includes a rich set of biological evidences on the anatomical
structure of the brain that may support his computation model.

The model uses HTMs (Hierarchical Temporal Memories) to build up intelligent devices. HTMs consist of
hierarchical tree-like P2P (Peer-To-Peer) networks where each node runs a similar {\em learning and
  predicting} algorithm. The adequacy of these HTMs to discover the hidden structure of our outside world lies
in the belief on the hierarchical structure of the world itself.

\begin{figure}\begin{center}
\includegraphics[scale=.40]{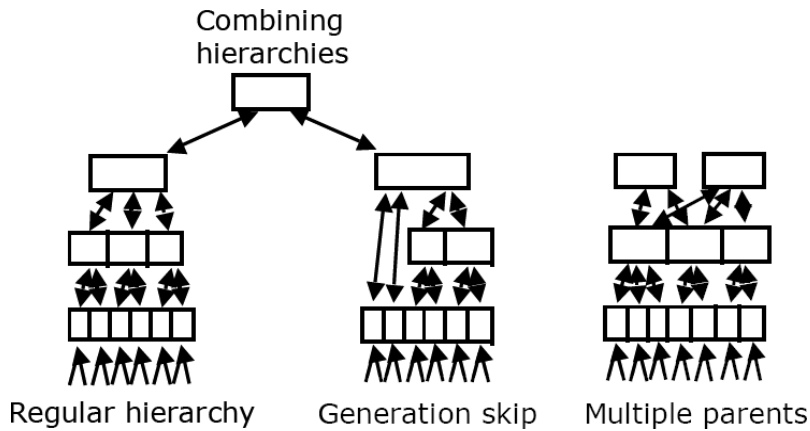}\vspace{-.5cm}\end{center}
\caption{Hierarchical Temporal Memories}\label{htm}\end{figure}

Fig.~\ref{htm} illustrates various types of HTMs. On the left, they are simple tree structures. On the right,
an example of an HTM with shared subtrees is given. This latter example shows that generally HTMs could have a
DAG (Directly Acyclic Graph) structure. However, as the processing flow may go up and down, or even between
nodes at the same level, the resulting graphs are pretty general. The hierarchical structure is useful mostly
as a conceptual approach for understanding the complicated structure and the complex activity of the brain.

\paragraph{Getting the right level of processing.}

Briefly, the activity of a mature brain is as follows. At each node, sequences of temporal patterns arrive and
are classified according to a learned schema. When this process is fully done, the patterns are replaced by
short codes of the classes they were classified into and the sequences of these codes are forwarded to an
upper node in the HTM hierarchy. This resembles the situation in a hierarchically structured company where an
employee tells his/her superior ``I have done this and this and this...,'' without entering into details.
During the classification process, a node may look at a few starting temporal data from its incoming pattern,
guesses the class the pattern falls into, and predicts the rest of the sequence. This gives a robust
processing in the case the input data are partially ambiguous or have errors.

Except for the explained forward flow of information in a HTM, there is also a feedback flow from upper to
lower nodes in the hierarchy. In the case a pattern can not be certainly classified by a node, the node might
forward the full pattern to his/her upper node in the HTM hierarchy asking for help. Using the above analogy,
this is like an employee telling his/her superior ``You see, I do not know what to do in this case. Could you
help me?''  Depending on the experience of the HTM (or of the brain that the HTM models), the right level of
processing the input patterns may be at a lower or a higher level in the hierarchy. Experienced HTMs/brains
tend to fully process the input patterns at lower levels, while inexperienced ones, for instance those found
in a learning process, tend to process the input patterns at higher levels in the hierarchy. Once a learned
process is stable at a hierarchy level, it can be shifted ``down to the hierarchy'' for latter
processing. This way, upper levels of the hierarchy become free and may process more abstract patterns,
concepts, or thoughts.

\paragraph{Associations.}

The focus in the previous paragraph was on the vertical structure in this HTM hierarchical model of the brain:
how to learn and classify the incoming pattern in isolation. Actually, the brain and the corresponding HTM
models have very powerful association mechanisms. These association mechanisms act either directly at a given
level in a hierarchy (nodes are informed on the activity of their close neighbors), or far-apart via the
hierarchical structure.  In the latter case, information from different sensory systems may be combined (e.g.,
simultaneous recognition of sounds and visual images) for a better and faster processing.

\paragraph{Perception and action.} 

While most of the intuition behind the above examples comes from the human perception system, this HTM model
of the brain makes no difference between the perception and the action mechanisms: the same kind of hierarchal
structure and the same processing mechanisms are present in the motor areas where brain thoughts are
translated into visible behaviors.

\paragraph{Numenta and intelligent machines.} 

The pitfalls of Hawkins' approach may come from the yet-to-be-discovered {\em learning and predicting
  algorithm} used in the nodes of the HTM models of the brain. While this might take long and might be very
difficult to discover, actually Hawkins has paved another way focusing on using HTMs to build ``intelligent
machines.'' His new company Numenta is planing to build computing chips based on HTM models and using
appropriate learning algorithms. Whether these algorithms do fit or not with the ones used by the brain may be
irrelevant - in design, we do not have to copy the nature: our cars have no legs, our planes have not
bird-like wings.

\paragraph{Turing test on intelligence.}

We close this brief presentation of Hawkins' model of the brain with a more philosophical discussion. What is
``intelligence'' and what it means for a computer to be ``as intelligent as a human being'' were (and still
are) long debated questions. Alan Turing has invented Turing machines, a mechanical model of computation on
which the modern computers are based. Turing has proposed this famous {\em Turing test}: a computer is as
intelligent as a human being if it is behaviorally equivalent with a human being. In other words, an external
observer can not see a difference between his/her interaction with a computer or with a human being.

Searle, a fervent critic of this kind of intelligence test, came up with a ``Chinese Room'' thought
experiment, showing that an English-speaking person following a set of rules (the analogy of a computer
program) can properly answer Chinese-written questions without actually understanding Chinese. His conclusion
is that intelligence and understanding can not be reduced to behavior.

Hawkins' model places more emphasis on ``prediction'' in his attempt to capture a definition for
intelligence. Understanding is closely linked with the capacity of prediction. Ultimately, understanding and
intelligence may be completely internal, in the brain, without any visible external behavior.

Ironically, the seminal paper of Turing contains a remark saying that what he has introduced is an {\em
  a-machine}, an autonomous one, and there is another notion of {\em s-machine}, an interactive one, which was
not considered there. This difference between a closed and an open (interactive) approach may explain the main
difference between Turing and Hawkins: Turing has used his autonomous machine reducing intelligence to its
external behavior, while Hawkins uses an interactive approach with a sophisticated dance between the
processing of the real input patterns and what the machine expects from its own prediction.

\section{Agapia - a parallel, interactive programming language}

Interactive computing \cite{gsv06} is a step forward on system modularization. The approach allows to describe
parts of the systems and to verify them in an open environment. A model for interactive computing systems
(consisting of interactive systems with registers and voices - {\em rv-systems}) and a core programming
language (for developing {\em rv-programs}) have been proposed in \cite{ste06a} based on register machines and
a space-time duality transformation. Structured programming techniques for rv-systems and a kernel programming
language Agapia have been later introduced \cite{dr-st07b}, with a particular emphasis on developing a
structural spatial programming discipline.

Structured process interaction greatly simplifies the construction and the analysis of interactive
programs. For instance, method invocation in current OO-programming techniques may produce unstructured
interaction patterns, with free {\tt goto}'s from a process to another and should be avoided. Compared with
other interaction or coordination calculi, the rv-systems approach paves the way towards a name-free calculus
and facilitates the development of a modular reasoning with good expectations for proof scalability to systems
with thousands of processes.  A new and key element in this structured interaction model is the extension of
temporal data types used on interaction interfaces. These new temporal data types (including voices as a
time-dual version of registers) may be implemented on top of streams as the usual data types are implemented
on top of Turing tapes.

Agapia \cite{dr-st07b,pss07} is a kernel high-level massively parallel programming language for interactive
computation. It can be seen as a coordination language on top of imperative or functional programming
languages as C++, Java, Scheme, etc.  Typical Agapia programs describe open processes located at various sites
and having their temporal windows of adequate reaction to the environment. The language naturally supports
process migration, structured interaction, and deployment of components on heterogeneous machines. Despite of
allowing these high-level features, the language can be given simple denotational and operational semantics
based on scenarios (scenarios are two-dimensional running patterns; they can be seen as the closure with
respect to a space-time duality transformation of the running paths used to define operational semantics of
sequential programs).

\subsection{Scenarios}\label{s-scen}

This subsection briefly presents temporal data, grids, scenarios, and operations on scenarios.

\paragraph{Temporal data.}

What we call ``spatial data'' are just the usual data occurring in imperative programming. For them, common
data structures and the usual memory representation may be used. On the other hand, ``temporal data'' is a
name we use for a new kind of (high-level) temporal data implemented on streams. A {\em stream} is a sequence
of data ordered in time. (The time model in Agapia is discrete.) Typically, a stream results by observing data
transmitted along a channel: it exhibits a datum (corresponding to the channel type) at each clock cycle.

A {\em voice} is defined as the time-dual of a register: {\em It is a temporal data structure that holds a
  natural number. It can be used (``heard'') at various locations. At each location it displays a particular
  value.}

Voices may be implemented on top of a stream in a similar way registers are implemented on top of a Turing
tape, for instance specifying their starting time and their length. Most of usual data structures have natural
temporal representations. Examples include timed booleans, timed integers, timed arrays of timed integers,
etc.

\paragraph{Grids and scenarios.}

\begin{figure}
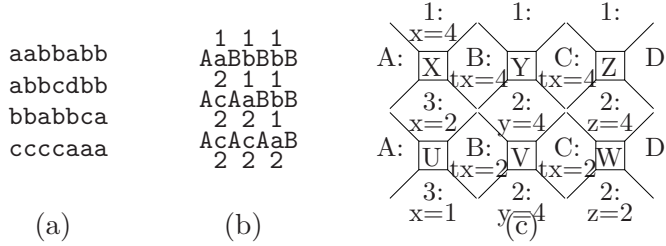
\begin{center}\hspace*{-.5cm}\begin{tabular}{c@{\hsp\hsp}c@{\hsp\hsp\hsp}c}
\raisebox{.7cm}{\tdwbis{small}{aabbabb\\abbcdbb\\bbabbca\\ccccaaa}}
& \raisebox{.9cm}{$\tdwbis{small}{\ 1\ 1\ 1\ \tdret AaBbBbB\tdret \ 2\ 1\
1\ \tdret AcAaBbB\tdret \ 2\ 2\ 1\ \tdret AcAcAaB\tdret 2\ 2\ 2\ }$}
&{\small$\begin{array}{@{}c@{}c@{}c@{}}
\scelli{1:\snvsp\\x=4}{A:}{X}{}{}
&\scelli{1:}{B:\snvsp\\tx=4}{Y}{}{}
&\scelli{1:}{C:\snvsp\\tx=4}{Z}{D}{}
\vspace{-1mm}\\
\scelli{3:\snvsp\\x=2}{A:}{U}{}{3:\snvsp\\x=1}
&\scelli{2:\snvsp\\y=4}{B:\snvsp\\tx=2}{V}{}{2:\snvsp\\y=4}
&\scelli{2:\snvsp\\z=4}{C:\snvsp\\tx=2}{W}{D}{2:\snvsp\\z=2}
\end{array}$}\\
(a)&(b)&(c)\end{tabular}\vspace{-.5cm}\end{center}
\caption{A grid (a), an abstract scenario (b), and a concrete scenario (c).\vspace{-.3cm}}\label{f-gscen}
\end{figure}

A {\em grid} is a {\em rectangular} two-dimensional array containing letters in a given alphabet.  A grid
example is presented in Fig.~\ref{f-gscen}(a). The default interpretation is that columns correspond to
processes, the top-to-bottom order describing their progress in time. The left-to-right order corresponds to
process interaction in a {\em nonblocking message passing discipline}: a process sends a message to the right,
then it resumes its execution.

A {\em scenario} is a grid enriched with data around each letter. The data may be given in an abstract form as
in Fig.~\ref{f-gscen}(b), or in a more detailed form as in Fig.~\ref{f-gscen}(c).

The type of a scenario interface is represented as $t_1;t_2;\dots;t_k$, where each $t_k$ is a tuple of simple
types used at the borders of scenario cells. The empty tuple is also written 0 or $nil$ and can be freely
inserted to or omitted form such descriptions. The type of a scenario is specified as $f:\tsrv{w}{n}{e}{s}$,
where $w,n,e,s$ are the types for its west, north, east, south interfaces.

\paragraph{Operations with scenarios.}

Two scenario interfaces $t=t_1;t_2;\dots;t_k$ and $t'=t'_1;t'_2;\dots;t'_{k'}$ are {\em equal}, written
$t=t'$, if $k=k'$ and the types and the values of each pair $t_i,t'_i$ are equal. Two interfaces are {\em
  equal up to the insertion of $nil$ elements}, written $t=_nt'$, if they become equal by appropriate
insertions of $nil$ elements.

\begin{figure}\begin{center}
\includegraphics[scale=.40]{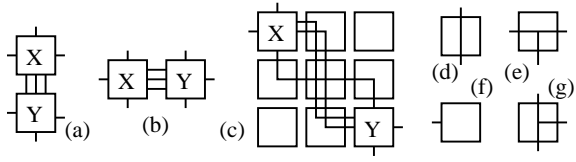}\vspace{-.5cm}\end{center}
\caption{Operations on scenarios}\label{f654}\end{figure}

Let $Id_{m,p}:\tsrv{m}{p}{m}{p}$ denote the constant cells whose temporal and spatial outputs are the same as
their temporal and spatial inputs, respectively; an example is the center cell in Fig.~\ref{f654}(c), namely
$Id_{1,2}$.

{\em Horizontal composition:} Let $f_i:\tsrv{w_i}{n_i}{e_i}{s_i}, i=1,2$ be two scenarios. Their {\em
  horizontal composition $f_1\hcomp f_2$} is defined only if $e_1=_nw_2$. For each inserted $nil$ element in
an interface (to make the interfaces $e_1$ and $w_2$ equal), a dummy row is inserted in the corresponding
scenario, resulting a scenario $\ol{f_i}$. The result $f_1\hcomp f_2$ is obtained putting $\ol{f_1}$ on left
of $\ol{f_2}$. The operation is briefly illustrated Fig.~\ref{f654}(b).  The result is unique up to insertion
or deletion of dummy rows. Its identities are $Id_{m,0}, m\geq 0$.

{\em Vertical composition:} The definition of {\em vertical composition $f_1\vcomp f_2$} (see
Fig.~\ref{f654}(a)) is similar, but now using the vertical dimension.  Its identities are $Id_{0,m}, m\geq 0$.

{\em Diagonal composition:} The {\em diagonal composition $f_1\dcomp f_2$} (see Fig.~\ref{f654}(c)) is a
derived operation defined only if $e_1=_nw_2$ and $s_1=_nn_2$. The result is defined by the formula\bi
\item[] $f_1\dcomp f_2 =(f_1\hcomp R_1\hcomp \Lambda)\vcomp(S_2\hcomp Id \hcomp R_2)\vcomp(\Lambda \hcomp
  S_1\hcomp f_2).$\ei for appropriate constants $R,S,Id,\Lambda$. Its identities are $Id_{m,n}, m,n\geq
  0$. (The involved constants $R,S,Id,\Lambda$ are described below.)

{\em Constants:} Except for the defined identities, we use a few more constants. Most of them can be found in
Fig.~\ref{f654}(c): a recorder $R$ (2nd cell in the 1st row), a speaker $S$ (1st cell in the 2nd row), an
empty cell $\Lambda$ (3rd cell in the 1st row). Other constants of interest are: transformed recorders
Fig.~\ref{f654}(e) and transformed speakers Fig.~\ref{f654}(g).

\subsection{Structured rv-programs}\label{s-srv}

\paragraph{The syntax of structured rv-programs.}

The basic blocks for constructing structured rv-programs are called {\em modules}. A module gets input data
from its west and north interfaces, process them (applying the module's code), and delivers the computed
outputs to its east and south interfaces.

On top of modules, structured rv-programs are built up using ``if'' and both, composition and iterated
composition statements for the vertical, the horizontal, and the diagonal directions. The composition
statements capture at the program level the corresponding operations on scenarios. The iteration statements
are also called the {\em temporal}, the {\em spatial}, and the {\em spatio-temporal while statements} - their
scenario meaning is described below.

The {\em syntax for structured rv-programs} is given by the following BNF grammar\bi\item[] $P::=
{}$\parbox[t]{8cm}{$X\ |\ if(C)then\{P\}else\{P\} |\ P\pvcomp P\ |\ P\phcomp P\ |\ P\pdcomp
  P\\ |\ while\_t(C)\{P\}\ |\ while\_s(C)\{P\} |\ while\_st(C)\{P\}$}\svsp\\ $X::=
{}$\parbox[t]{8cm}{$module\{listen\ t\_vars\}\{read\ s\_vars\}\\\{code;\}\{speak\ t\_vars\}\{write\ s\_vars\}$}\ei
This is a core definition of structured rv-programs, as no data types or language for module's code are
specified. Agapia, to be shortly presented, is a concrete incarnation of structured rv-programs into a fully
running environment.

Notice that we use a different notation for the composition operators on scenarios $\vcomp,\hcomp,\dcomp$ and
on programs $\pvcomp,\phcomp,\pdcomp$; moreover, the extension of the usual composition operator ';' to
structured rv-programs is denoted by ``$\pvcomp$''.

\paragraph{Operational semantics.}

The operational semantics \snvsp$$|\ \ |:\mbox{Structured rv-programs}\ra \mbox{Scenarios}\snvsp$$ associates
to each program the set of its running scenarios.

The type of a program $P$ is denoted $P:\tsrv{w(P)}{n(P)}{e(P)}{s(P)}$, where $w(P)/n(P)/e(P)/s(P)$ indicate
its types at the west/north/east/south borders. On each border, the type may be quite complex - the convention
is to separate by ``,'' the data from within a module and by ``;'' the data coming from different
modules. This convention applies to both spatial and temporal data.

We say, two interface types {\em match} if they have a nonempty intersection.

\paragraph{Modules.} 

The modules are the starting blocks for building structured rv-programs. The {\tt listen (read)} instruction
is used to get the temporal (spatial) input and the {\tt speak (write)} instruction to return the temporal
(spatial) output. The {\tt code} consists of simple instructions as in the C code. No distinction between
temporal and spatial variables is made within a module.

A scenario for a module consists of a unique cell, with concrete data on the borders, and such that the output
data are obtained from the input data applying the module's code.

\paragraph{Composition.} 

\begin{figure}\begin{center}\begin{tabular}{c@{\hspace{1cm}}c@{\hspace{1cm}}c}
\includegraphics[scale=.4]{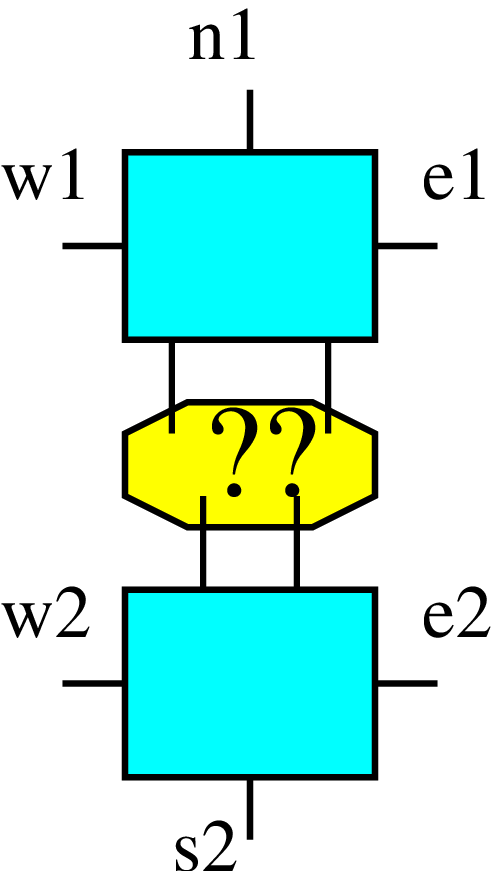} & \raisebox{.75cm}{\includegraphics[scale=.4]{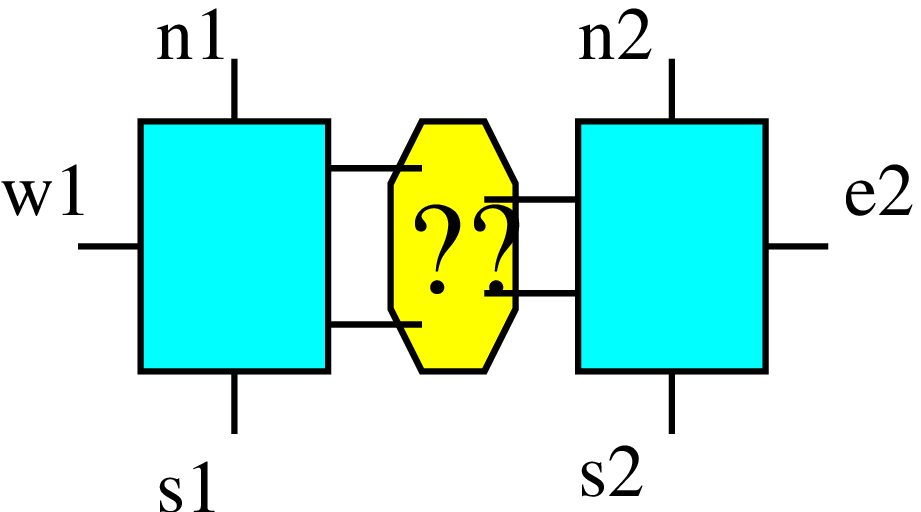}}
& \raisebox{.4cm}{\includegraphics[scale=.4]{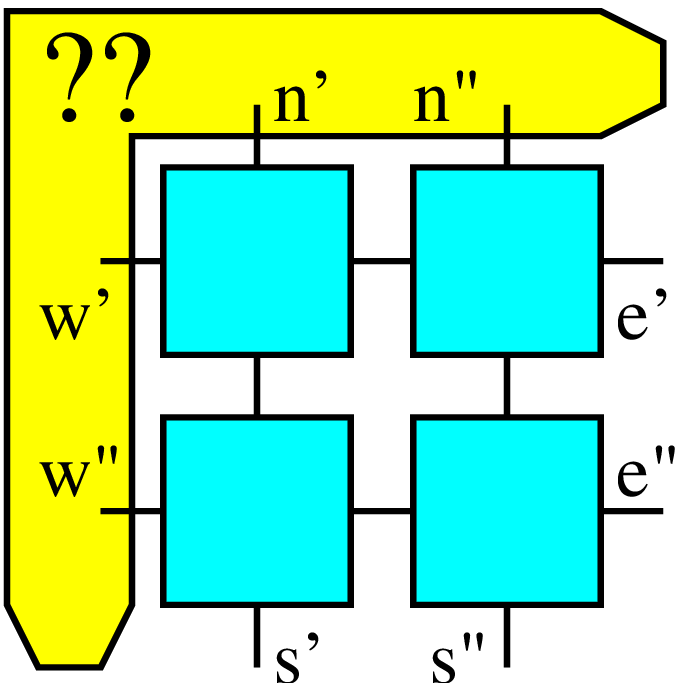}}\vspace{-.75cm}
\end{tabular}\end{center}\caption{The vertical and the horizontal compositions and the ``if'' statement}
\label{vh+if}\end{figure}

Programs may be composed ``horizontally'' and ``vertically'' as long as their types on the connecting
interfaces agree. They can also be composed ``diagonally'' by mixing the horizontal and vertical compositions.

For two programs $P_i:\tsrv{w_i}{n_i}{e_i}{s_i}$, $i=1,2$ we define the following composition operators.

{\em Horizontal composition:} $P_1\phcomp P_2$ is defined if the interfaces $e_1$ and $w_2$ match, see
Fig.~\ref{vh+if}(middle). The type of the composite is $\tsrv{w_1}{n_1;n_2}{e_2}{s_1;s_2}$. A scenario for
$P_1\phcomp P_2$ is a horizontal composition of a scenario in $P_1$ and a scenario in $P_2$.

{\em Vertical composition:} $P_1\pvcomp P_2$ is similar.

{\em Diagonal composition:} $P_1\pdcomp P_2$ is defined if $e_1$ matches $w_2$ and $s_1$ matches $n_2$. The
type of the composite is $\tsrv{w_1}{n_1}{e_2}{s_2}$. A scenario for $P_1\pdcomp P_2$ is a diagonal
composition of a scenario in $P_1$ and a scenario in $P_2$.

\paragraph{If.}

For two programs $P_i:\tsrv{w_i}{n_i}{e_i}{s_i}$, $i=1,2$, a new program
$Q=if\ (C)\ then\ \{P_1\}\ else\ \{P_2\}$ is constructed, where $C$ is a condition involving both, the
temporal variables in $w_1\cap w_2$ and the spatial variables in $n_1\cap n_2$, see
Fig.~\ref{vh+if}(right). The type of the result is $Q:\tsrv{w_1\cup w_2}{n_1\cup n_2}{e_1\cup e_2}{s_1\cup
  s_2}$.

A scenario for $Q$ is a scenario of $P_1$ when the data on west and north borders of the scenario satisfy
condition $C$, otherwise is a scenario of $P_2$ (with these data on these borders).

\paragraph{While.} 

Three types of while statements are used for defining structured rv-programs, each being the iteration of a
corresponding composition operation.

{\em Temporal while:} For $P:\tsrv{w}{n}{e}{s}$, the statement $while\_t\ (C)\{P\}$ is defined if the
interfaces $n$ and $s$ match and $C$ is a condition on the spatial variables in $n\cap s$. The type of the
result is $\tsrv{(w;)^*}{n\cup s}{(e;)^*}{n\cup s}$. A scenario for $while\_t\ (C)\{P\}$ is either an
identity, or a repeated vertical composition $f_1\vcomp f_2\vcomp\dots\vcomp f_k$ of scenarios for $P$ such
that: (1) the north border of each $f_i$ satisfies $C$ and (2) the south border of $f_k$ does not satisfy $C$.

{\em Spatial while:} $while\_s\ (C)\{P\}$ is similar.

{\em Spatio-temporal while:} $while\_st\ (C)\{P\}$, where $P:\tsrv{w}{n}{e}{s}$, is defined if $w$ matches $e$
and $n$ matches $s$ and, moreover, $C$ is a condition on the temporal variables in $w\cap e$ and the spatial
variables in $n\cap s$. The type of the result is $\tsrv{w\cup e}{n\cup s}{w\cup e}{n\cup s}$. A scenario for
$while\_st\ (C)\{P\}$ is either an identity, or a repeated diagonal composition $f_1\dcomp
f_2\dcomp\dots\dcomp f_k$ of scenarios for $P$ such that: (1) the west and north border of each $f_i$
satisfies $C$ and (2) the east and south border of $f_k$ does not satisfy $C$.

\subsection{Agapia}

\paragraph{Syntax of Agapia v0.1 programming language.}

\begin{figure}[t]\parbox[t]{7.3cm}{\footnotesize
{\bf Interfaces}\svsp\\
$SST::={}$ \parbox[t]{7cm}{$nil\ |\ sn\ |\ sb\ |\ (SST\cup SST)\\ |\ (SST,SST)\ |\
(SST)^*$}\svsp\\
$ST::={}$ \parbox[t]{7cm}{$ (SST)\ |\ (ST\cup ST)\\ |\ (ST;ST)\ |\ (ST;)^*$}\svsp\\
$STT::={}$ \parbox[t]{7cm}{$nil\ |\ tn\ |\ tb\ |\ (STT\cup STT)\\ |\ (STT,STT)\ |\
(STT)^*$}\svsp\\
$TT::= {}$ \parbox[t]{7cm}{$(STT)\ |\ (TT\cup TT)\\ |\ (TT;TT)\ |\ (TT;)^*$}\vsp

{\bf Expressions}\svsp\\
$V::={}$ \parbox[t]{7cm}{$ x:ST\ |\ x:TT\ |\ V(k)\\ |\ V.k\ |\ V.[k]\ |\ V@k\ |\ V@[k]$}\svsp\\
}\hspace*{-1.5cm}\parbox[t]{9cm}{\footnotesize
$E::= n\ |\ V\ |\ E+E\ |\ E*E\ |\ E-E\ |\ E/E$\svsp\\
$B::= b\ |\ V\ |\ B\&\& B\ |\ B||B\ |\ !B\ |\ E<E$\vsp\\
{\bf Programs}\svsp\\
$W::={}$ \parbox[t]{7cm}{$nil\ |\ new\ x:SST\ |\ new\ x:STT\\ |\ x := E\ |\ if (B) \{W\} else \{W\}
\\ |\ W;W\ |\ while (B) \{W\}$}\svsp\\
$M::={}$ \parbox[t]{7cm}{$module\{listen\ x:STT\}\{read\ x:SST\}\\
\{W;\}\{speak\ x:STT\}\{write\ x:SST\}$}\svsp\\
$P::={}$ \parbox[t]{7cm}{$nil\ |\ M\ |\ if (B) \{P\} else \{P\}\ \\ 
|\ P\pvcomp P\ |\ P\phcomp P\ |\ P\pdcomp P\ \\
|\ while\_t (B) \{P\}\ |\ while\_s (B) \{P\}\\ |\ while\_st (B) \{P\}$}}
\caption{The syntax of Agapia v0.1 programs}\label{f-agapia}
\end{figure}

The syntax for Agapia v0.1 programs is presented in Fig.~\ref{f-agapia}. The v0.1 version is intentionally
kept simple to illustrate the key features of the approach (see \cite{pss07} for an extension v0.2 including
high-level structured rv-programs). Agapia v0.1 forms a kind of minimal interactive programming languages: it
describes what can be obtained from classical while programs allowing for spatial and temporal integer and
boolean types and closing everything with respect to space-time duality.

The types for spatial interfaces are built up starting with integer and boolean $sn,sb$ types, applying the
rules for $\cup,\mbox{','},(\_)^*$ to get process interfaces, then the rules for $\cup,\mbox{';'},(\_;)^*$ to
get system interfaces. The temporal types are similarly introduced. For a spatial or temporal type $V$, the
notations $V(k),V.k,V.[k], V@k,V@[k]$ are used to access its components. Expressions, usual while programs,
modules, and programs are then naturally introduced.

\section{Agapia programs for HTMs models}

The current approach is to give Agapia scenario-based semantics with linear models for space and time. When
different models are needed, as tree models for the HTMs presented in this paper, a linear representation of
such models is required. Fortunately, there is a huge amount of work on similar topics involving
representation of an endless number of data structures in the linear virtual memory model of conventional
computers.

We focus our design of Agapia programs below on the HTM in the left side of Fig.~\ref{htm} (the regular
hierarchy, restricted to 2 levels: top, level 1, level 2) considered as a HTM model for a part of a visual
sensory system. 

Tree structures are represented recursively labeling their nodes by strings as follows: ``if a node $p$ in the
tree is labelled by $w$ and a node $q$ in the tree is his/her $i$-th son (direct descendent), counting the
positions from left to right, then the code of $q$ is $wi$.''

In our example, the codes of the nodes are: nil (for the top node), 1,2,3 (for the nodes on level 1), and
11,12,21,22,31,32 (for the nodes on level 2).  The nodes are placed in a linear order using an extension of
the Left-Right-Root parsing in binary trees. In our case the result is: 11,12,1,21,22,2,31,32,3,nil. With this
convention, it is easier to describe Agapia programs for the forward flow of information, but slightly more
complicated for the feedback flow. In the sequel, we suppose each process knows his/her code in the list and
the codes of other nodes in the structure.

Our approach to modeling HTMs with Agapia programs consists of three steps. 

The first step requires more or less conventional programming for the basic modules. As one can develop Agapia
on top of C++ or Java, code in such languages could be used in these modules. The code has to implement the
following features.\be

\item Each node has a classifying algorithm, for instance using a ``typical representative'' for each class.
  Suppose the classes are $\{C_1,\dots,C_n\}$ and their representatives are $\{t_1,\dots,t_n\}$. Given a
  temporal pattern $t$ (a voice, or a more complex temporal data structure) the node find the best matching of
  $t$ against $t_1,\dots,t_n$. According to Hawkins, this classification should be unique. However,
  increasingly sophisticated procedures may be used to reach this result, for instance using the feedback flow
  from top-to-bottom in the hierarchy or the association flow from the neighboring nodes.

\item An alternative to the {\em best full matching} is to use a {\em prefix matching}: once a prefix of $t$
  was parsed and it fits with a unique $t_i$, then the rest of $t$ is ignored.

\item Each class $C_i$ is supposed to have a name $n_i$ (codified with much shorter sequences than for $t$ or
  $t_i$'s). The final product of the node is the passing of the code $n_k$ of the class $C_k$ for which $t$
  has the best fit to his/her HTM parent.

\item In contrast with 2, another alternative is to have {\em fully attentive} nodes, keeping track on all
  details. In such a case, if the input pattern does fit well with none of $t_i$'s, the node passes the full
  $t$ (not just a code for his/her better fitting class) to his/her parent for further processing.

\item Finally, higher level nodes, except for their own classification, have to process the exceptions in the
  classification procedures of their descendent nodes. \ee

The second step is to describe the forward flow of information in this HTM. It is just a particular format of
MPI-like communication mechanisms for which Agapia macro-programs can be easily written (see, e.g.,
\cite{ch-st08a}). The shape of the program is as follows.\be
 
\item The program contains a main diagonal while statement. It repeats the processing for repeated incoming
  temporal patterns $t$'s.

\item During a step of the diagonal while statement, each node gets input patterns from the left, processes
  them, and passes the results to the right. This means, for the leaves the input data come from outside (from
  the open temporal interfaces), while for the inner nodes they come from their own descendents. The results
  are passed to the parents, which fortunately are placed on the right in this linear order of the nodes. This
  way, when a body of the diagonal while statement is executed, the forward flow in the HTM is fully
  modeled.\ee

The third step is to show how the feedback flow in the HTM can be modeled. This is slightly more complicated,
as we have chosen a linear order to facilitate the modeling of the forward flow. Indeed, as the interaction in
Agapia programs goes from left to right, when a parent node wants to send a message to a son, an extra
diagonal composition is needed to model this communication. Except for the extra diagonal composition steps,
the modeling of the interaction is similar to the previous case.

\section{TRIPS - a multi-core/many-core architecture}

With a privileged role between programming languages and hardware, the instruction set used in computer
architecture design is very conservative. Changing or introducing new {\em ISA (Instruction Set Architecture)}
is disruptive for computer systems and may be very risky. Nevertheless, the time for a radical change is
imperative. The old CISC/RISC instruction sets no longer fit with the huge potential of the forthcoming
multi-core/many-core computers. Introducing new ISA is now worthwhile having the potential to address the
challenges of modern technologies and to exploit various integration possibilities \cite{computer}. In this
context, {\em TRIPS (Tera-op, Reliable, Intelligently adaptive Processing System)} architectures are a very
promising recent proposal facilitating higher exploitation of data, instruction-level, and thread-level
parallelisms \cite{url-trips}.

TRIPS is an instantiation of the {\em EDGE (Explicit Data Graph Execution)} ISA concept. EDGE is a new class
of ISAs that views an instruction stream as blocks of instructions for a single task using isolated data. The
main feature of an EDGE architecture refers to direct instruction communication which enables a dataflow-like
execution. Unlike RISC and CISC instruction sets, EDGE explicitly encodes dependences into individual
instructions. The hardware is not required to rediscover data dependences dynamically at runtime because the
compile-time dependence graph is expressed through the ISA. Higher exposed concurrency and power-efficient
execution are therefore facilitated by an EDGE architecture \cite{computer}. EDGE overcomes major drawback
issues of the RISC and CISC architectures such as the usage of inefficient and power-consuming structures.
 
Offering increased flexibility, TRIPS supports a static placement of instructions (driven by compiler) and
dynamic issue (hardware-determined) execution model. Graphs of predicated hyperblocks are compiled and
represented internally as a dataflow graph. Communication between hyperblocks is possible via a set of input
and output registers.

The TRIPS architecture aims to increase concurrency, to achieve power-efficient high performance and to
diminish communication delays.  Concurrency is increased by using an array of arithmetic logic units (ALUs)
executed concurrently. ALUs provide scalable issue width as well as scalable instruction window size
\cite{computer}. The TRIPS architecture minimizes execution delays by using compile-time instruction
placement. Computation patterns are efficiently supported by the dataflow-like execution model of TRIPS.

The block-atomic execution engaged in TRIPS works as follows \cite{computer}:
\begin{itemize}
\item Instructions are grouped by the compiler into groups of instructions called hyperblocks (each hyperblock
  contains up to 128 instructions) and mapped to an array of execution units;
\item Each hyperblock is fetched, executed, and committed {\em atomically}; instructions are fetched in
  parallel and loaded into the instruction buffers at each ALU;
\item Instructions are efficiently executed by the hardware using a fine-grained dataflow model.
\end{itemize}

TRIPS can effectively support parallelism (instruction-level, data-level, and thread-level parallelism). As
long as the software can discover parallelism, the TRIPS architecture will effectively exploit it.  Being easy
to scale up and down in performance, TRIPS overcomes the scheduling problems of traditional designs as well as
the explicit unit exposure of VLIW (Very Long Instruction Word) designs.

\section{Running Agapia programs on TRIPS architectures}

We end our trip from brain models to multi-core/many-core computers with some remarks on the possibility of
compiling and running Agapia programs on TRIPS architectures.

\begin{figure}[tbh]
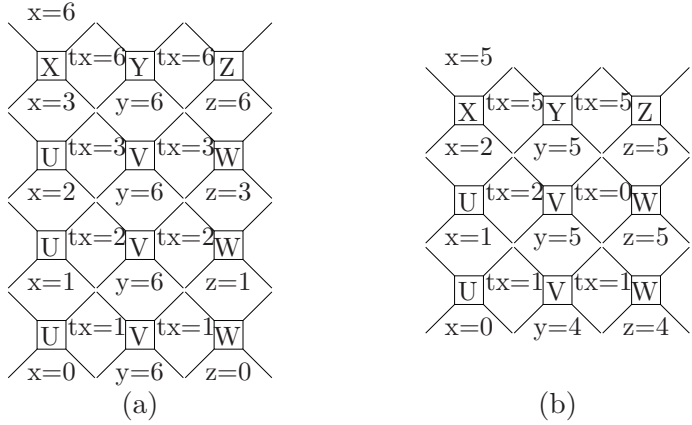
\begin{center}\begin{tabular}{c@{\hspace{2cm}}c}
{\small$\begin{array}{@{}c@{}c@{}c@{}}
\scelli{x=6}{}{X}{}{}&\scelli{}{tx=6}{Y}{}{}&\scelli{}{tx=6}{Z}{}{}\vspace{-1mm}\\
\scelli{x=3}{}{U}{}{}&\scelli{y=6}{tx=3}{V}{}{}&\scelli{z=6}{tx=3}{W}{}{}\vspace{-1mm}\\
\scelli{x=2}{}{U}{}{}&\scelli{y=6}{tx=2}{V}{}{}&\scelli{z=3}{tx=2}{W}{}{}\vspace{-1mm}\\
\scelli{x=1}{}{U}{}{x=0}&\scelli{y=6}{tx=1}{V}{}{y=6}&\scelli{z=1}{tx=1}{W}{}{z=0}\end{array}$}
&{\small$\begin{array}{@{}c@{}c@{}c@{}}
\scelli{x=5}{}{X}{}{}&\scelli{}{tx=5}{Y}{}{}&\scelli{}{tx=5}{Z}{}{}\vspace{-1mm}\\
\scelli{x=2}{}{U}{}{}&\scelli{y=5}{tx=2}{V}{}{}&\scelli{z=5}{tx=0}{W}{}{}\vspace{-1mm}\\
\scelli{x=1}{}{U}{}{x=0}&\scelli{y=5}{tx=1}{V}{}{y=4}&\scelli{z=5}{tx=1}{W}{}{z=4}\end{array}$}\\
(a)&(b)\end{tabular}\vspace{-.3cm}\end{center}
\caption{Scenarios for perfect numbers}\label{s-perfect}\end{figure}

To illustrate the approach, we consider a simple example involving perfect numbers. A number is perfect if it
is equal to the sum of its proper divisors. Before showing Agapia programs for perfect numbers, we describe
two typical running scenarios for this task (one for a perfect number, the other for an imperfect one) in
Fig.~\ref{s-perfect}. The input-output relation is: {\it if the input number in the upper-left corner is $n$,
  then the output number in the lower-right corner is 0 iff $n$ is perfect}.

The scenarios in Fig.~\ref{s-perfect} use cells whose behaviors are captured by the modules in
Table~\ref{t-mod}.

\begin{table}[tbh]\rule{\textwidth}{.1mm}\nvsp\bi\item[]{\small\tt
module X\{listen nil;\}\{read x:sn;\}\\
  \hspace*{1cm}\{tx:tn; tx=x; x=x/2;\}\{speak tx;\}\{write x;\}\\
module Y\{listen tx:tn;\}\{read nil;\}\\
  \hspace*{1cm}\{y:sn; y=tx;\}\{speak tx;\}\{write y;\}\\
module Z\{listen tx:tn;\}\{read nil;\}\\
  \hspace*{1cm}\{z:sn; z=tx;\}\{speak nil;\}\{write z;\}\\
module U\{listen nil;\}\{read x:sn;\}\\
  \hspace*{1cm}\{tx:tn; tx=x; x=x-1;\}\{speak tx;\}\{write x;\}\\
module V\{listen tx:tn;\}\{read y:sn;\}\\
  \hspace*{1cm}\{if(y\%tx != 0) tx=0;\}\{speak tx;\}\{write y;\}\\
module W\{listen tx:tn;\}\{read z:sn\}\\
  \hspace*{1cm}\{z=z-tx;\}\{speak nil;\}\{write z;\}\\
module U1\{listen nil;\}\{read x:sn;\}\\
  \hspace*{1cm}\{tx:tn; tx=-1;\}\{speak tx;\}\{write nil;\}\\
module V1\{listen tx:tn;\}\{read y:sn;\}\\
  \hspace*{1cm}\{null;\}\{speak tx;\}\{write nil;\}\\
module W1\{listen tx:tn;\}\{read z:sn\}\\
  \hspace*{1cm}\{null;\}\{speak nil;\}\{write z;\}}\nvsp\ei\rule{\textwidth}{.1mm}
\caption{Modules for ``perfect numbers'' programs}\label{t-mod}\end{table}

Our first Agapia program \textbf{Perfect1} corresponds to the construction of the scenarios by rows:
{\tt\bfseries\snvsp\bi
\item[](X\ \phcomp\ Y\ \phcomp\ Z)\ \pvcomp\ while\_t(x > 0)\{U\ \phcomp\ V\ \phcomp\ W\}\snvsp\ei} The type
of the program is $\textbf{Perfect1}:\tsrv{nil}{sn;nil;nil}{nil}{sn;sn;sn}$.  Actually, the result is a
program similar with a usual imperative program. There are some ``transactions,'' each transaction specifying
a macro-step in the whole system. The interaction part is simple and it reduces to the interaction of the
components in a macro-step.

Our second Agapia program \textbf{Perfect2} corresponds to the construction of the scenarios by
columns:{\tt\bfseries\snvsp\bi\item[] (X\ \pvcomp\ while\_t(x >
  0)\{U\}\ \pvcomp\ U1)\\ \ \phcomp\ (Y\ \pvcomp\ while\_t(tx >
  -1)\{V\}\ \pvcomp\ V1)\\ \ \phcomp\ (Z\ \pvcomp\ while\_t(tx > -1)\{W\}\ \pvcomp\ W1)\snvsp\ei} Its type is
$\textbf{Perfect2}:\tsrv{nil}{sn;nil;nil}{nil}{nil;nil;sn}$.  This variant resembles the dataflow computation
paradigm. Each component acts as a stream processing function and the overall result comes from the
interaction of these components.

The first program is appropriate for running on classical architectures, while the last one for dataflow
architectures. TRIPS architecture is a combination of both. The current prototype uses sequences of up to 128
instructions to feed its matrix of ALUs. Agapia is very flexible and expressive, for instance the above two
programs are just the extreme cases of a rich variety of possibilities. More precisely, one could unroll the
first program, or restrict the number of steps in each component in the second program to get programs which
fit better with the TRIPS architecture. Such transformations might be performed automatically to help the user
to focus on the logic of the program and not on the target computer running his/her program.

Compiling Agapia programs for TRIPS architecture is without a doubt a very challenging direction. While our
intuition strongly supports such an attempt, the painful procedure of writing a compiler and running programs
is actually needed to clarify how well Agapia language and TRIPS architecture fit together.

\section{Conclusions and future work}

Most programs for AI tasks are inefficient on traditional computers with their Von-Neumann architecture and
using imperative programming style. The proposed dataflow machines from eighties and nineties, specific for AI
tasks, never had a major impact on the market. What we see in the recently proposed TRIPS architectures is a
combination of dataflow and Von-Neumann styles, particularly using speculative execution of long blocks of
instructions on the computers' arrays of ALUs.

The speculation on possible executions of the paths in a program, used to increase the processing speed, looks
somehow similar to the prediction process in the HTM models of the brain. A comparison between these two
computing models may be worthwhile for both fields. For instance, while the computer prediction is in the
narrow window of the user program demand, the HTM models of the brain are more open, interactive, agent-like -
here the prediction is mixed with possible actions of the human being who can change the course of the
forthcoming input data. This may explain why humans are good on interactive tasks, while current computers
with their predefined program-captured behavior are not.

We plan to develop the ideas from this paper in both directions: (1) to get a rich set of Agapia programs for
HTMs models, particularly for those used by Numenta platform \cite{url-numenta}; and (2) to explore the
possibility of getting a running environment for Agapia programs on TRIPS computers \cite{url-trips}.

\paragraph{Acknowledgments.} 

Part of Camelia Chira's work on this paper was carried out during a visit to the Department of Computer
Science of the University of Illinois at Urbana-Champaign, which was supported by a grant from CNCSIS,
Romania.  Gheorghe Stefanescu would like to thank Mircea Stan for pointers to TRIPS architectures.

\end{document}